\documentclass[amsmath,amssymb]{revtex4-1}

\usepackage{graphicx}
\usepackage{bm}

\def\be{\begin{equation}}
\def\ee{\end{equation}}
\def\bee{\begin{eqnarray}}
\def\eee{\end{eqnarray}}

\begin{document}

\preprint{Preprint: do not distribute}

\title{The linear tearing instability in three dimensional, toroidal gyrokinetic simulations.}
\author{W.A.~Hornsby, P.~Migliano, R.~Bucholz, L.~Kroenert, A.G.~Peeters}
\affiliation{Theoretical Physics V, Dept. of Physics, Universitaet Bayreuth, Bayreuth, Germany, D-95447}
\email{william.hornsby@ipp.mpg.de}

\author{D.Zarzoso, E.~Poli}
\affiliation{Max-Planck-Institut f\" ur Plasmaphysik, Boltzmannstrasse 2, D-85748
Garching bei M\" unchen, Germany} 

\author{F.J.~Casson}
\affiliation{CCFE, Culham Science Centre, Abingdon, \\
Oxon, OX14 3DB, UK} 

\date{\today}

\begin{abstract}

Linear gyro-kinetic simulations of the classical tearing mode in three-dimensional toroidal geometry  were performed
using the global gyro kinetic turbulence code, GKW .  The results were benchmarked against a cylindrical ideal MHD and analytical theory calculations. The stability, growth rate and frequency of the mode were investigated by varying the current profile, collisionality and the pressure gradients.   Both collision-less and semi-collisional tearing modes were found with a smooth transition between the two.  A residual, finite, rotation frequency of the mode even in the absense of a pressure gradient is observed which is attributed to toroidal finite Larmor-radius effects.  When a pressure gradient is present at low collisionality, the mode rotates at the expected electron diamagnetic frequency.  However the island rotation reverses direction at high collisionality.   The growth rate is found to follow a $\eta^{1/7}$ scaling with collisional resistivity in the semi-collisional regime, closely following the semi-collisional scaling found by Fitzpatrick.  The stability of the mode closely follows the stability using resistive MHD theory, however a modification due to toroidal coupling and pressure effects is seen. 

\end{abstract}

\maketitle

\section{Introduction}

A current density profile in a plasma can lead to instabilities that are able to change the topology
of the magnetic field via the process of magnetic reconnection \cite{Bisk}. Resistive instabilities, and in particular tearing instabilities \cite{NEW60,FUR63,FUR73} are found in a wide variety of astrophysical and laboratory situations \cite{Zwei09}. 

In a magnetically confined tokamak plasma, reconnecting instabilities can form magnetic islands.  This has a detrimental effect on the plasma confinement caused by the introduction of a radial component of the magnetic field \cite{WAE09}.   The tearing mode can trigger a disruption, something which would be catastrophic to the ITER  tokamak and therefore the tearing mode, and specifically, the neoclassical tearing mode (NTM) \cite{CAR86,HEG98} is expected to be the limiting factor on its operational plasma $\beta$ \cite{SAU97}.

Much work has been performed numerically in studying tearing mode stability in both cylindrical and toroidal geometry \cite{NISH98}, using the Magnetohydrodynamic (MHD) framework.  However for large current
and future tokamaks a fully kinetic description of the physics in the resonant layer is needed for both the linear and non-linear regimes.   

\begin{figure*}
\centering
\includegraphics[width=16.0cm,clip]{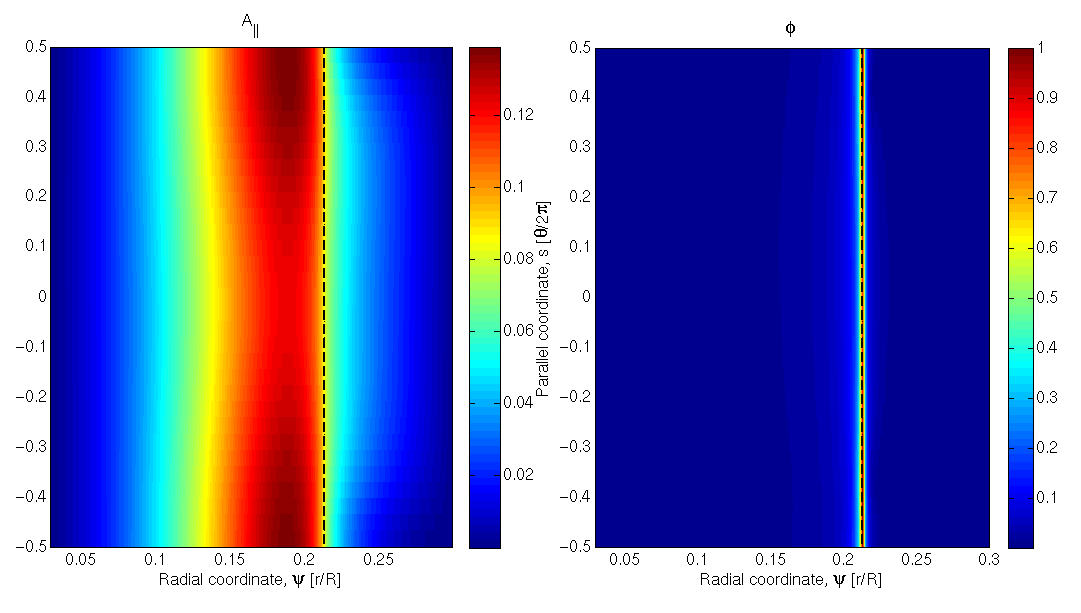}
\caption{Contour plots of (left) the parallel-radial mode structure of $A_\parallel$ and (right) the 
electrostatic potential, $\phi$ for a $m=2$,$n=1$ tearing mode.  The vertical dashed line denotes the 
position of the $q=2$ rational surface.}  
\label{ModeStruct}
\end{figure*}

Linear theory is generally considered to be valid when the generated magnetic island has a width that is smaller than the singular layer.  Above this width, nonlinear effects slow the growth of the island to an algebraic scaling \cite{RUTH73} and eventually cause the island to saturate \cite{Has04}.  Recent work studying the effects of electromagnetic turbulence on the evolution of the tearing mode
indicates a turbulent modification of the resonant layer, extending the suitability of linear theory to larger island sizes\cite{PRL14}, while it has also been seen that a magnetic island can have a profound effect on turbulence and transport \cite{POL09,HorEPL}.  

Within the singular layer (Parallel wave-vector of the mode, $k_{||}=0$)  the  assumptions of ideal MHD break down and magnetic reconnection may take place.  For a collisionless mode the mechanism of reconnection is the electron inertia, where the singular layer width, and in turn the growth rate, are related closely to the electron skin depth \cite{HAZ75}. When collisions become significant, it is plasma resistivity which determines the singular layer width, producing a classical collisional tearing \cite{FUR63} or, more relevant for present day tokamaks, a semi-collisional mode \cite{DRA77,Fitz10}.  The mode frequency, and in particular, its direction is vital in predicting the nonlinear stabity of the NTM.  Specifically the sign of the mode frequency, whether the mode rotates in the ion or electron diamagnetic direction,  determines whether the polarisation current has a stabilising or destabilising effect \cite{WIL96,Fitz06} particularly when the magnetic island is small \cite{Pol05}.

In this work the linear mode stability, growth rate and frequency is studied for the first time in three-dimensional toroidal geometry using the gyro-kinetic framework of equations.  Parameters appropriate to modern and future Tokamak experiments are used.  Gyrokinetics has been highly successful when applied to the study of drift waves, turbulence and transport but its use in the study of large scale MHD instabilities is still in its infancy.  Recent work has studied two dimensional magnetic reconnection in a highly magnetised system \cite{Wan05,NUM09,ROG07} and the ideal kink instability \cite{Mish}.

The paper is organised as follows.  In Sec.\ref{Framework} we describe the equations solved, while in Sec.~\ref{Implementation} we outline how the tearing mode is implemented within the gyrokinetic framework.  Sec.~\ref{Parameters} outlines the parameters used, and the resolution requirements.  Sec.~\ref{curprof} and \ref{colleff} presents the scaling of the tearing mode growth rate and mode frequency with collisionality, and with the variation of the background density, temperature and current profiles. 

\section{Gyrokinetic framework}
\label{Framework}

The self-consistent treatment of the tearing mode requires a radial profile in geometry and thermodynamic quantities.  We use here the global version of the gyrokinetic code, GKW \cite{PEEGlo}.  This is advancement on the already widely used flux-tube variant \cite{PEE09}.

The code solves the gyrokinetic set of equations.  The full details can
be found in \cite{PEE09} and references found therein.  Here we outline the basic set of equations that are solved and in the next section the modification and assumptions used in driving the tearing instability.

The delta-$f$ approximation is used.  The distribution function is split into a background $F$ and a perturbed distribution $f$.
The final equation for the perturbed distribution function $f$, for each species, $s$, can be written in the form 
\be 
{\partial g \over \partial t} + (v_\parallel {\bf b} + {\bf v}_D) \cdot \nabla f   
-{\mu B \over m}{{\bf B}\cdot \nabla B \over B^2}{\partial f \over \partial v_\parallel} = S, 
\label{gyrovlas}
\ee
where $S$ is the source term which is determined by the analytically known, background distribution function, $\mu$ is the magnetic moment, $v_{||}$ is the velocity along the magnetic field, $B$ is the magnetic field strength, m and Z are the particle mass and charge number respectively. Here,  $g = f + (Ze/T)v_{\parallel}\langle A_{\parallel} \rangle F_{M}$ is used to absorb the time derivative of the parallel vector potential $\partial A_{\parallel}/\partial t$ which enters the equations through Amp\`{e}res law.  

The velocities in Eq.~(\ref{gyrovlas}) are from left to right: the parallel motion along the 
unperturbed field ($v_\parallel {\bf b}$), the drift motion due to the inhomogeneous field 
(${\bf v}_D =  {1\over Ze} \biggl [ {m v_\parallel^2\over B} + \mu \biggr ] {{\bf B} \times \nabla B \over B^2}$), and the motion due to the perturbed electromagnetic field (${\bf v}_\chi  = {{\bf b} \times \nabla \chi \over B}$, where $\chi = \langle\phi\rangle - v_{||}\langle A_{||}\rangle$).   The latter is the combination of the $E \times B$ velocity (${\bf v}_E = {\bf b} \times \nabla \langle \phi \rangle / B$) and the parallel motion along 
the perturbed field line (${\bf v}_{\delta B} = - {{\bf b} \times \nabla v_\parallel 
\langle A_\parallel \rangle /B}$).  Here, the angled brackets denote gyro-averaged quantities.

The background is assumed to be a shifted Maxwellian ($F_M$), with particle density ($n(r)$) and temperature ($T(r)$) and toroidal flow velocity ($\omega_\phi(r)$)
\be 
\label{maxwell}
F_{Ms} = {n_{s} \over \pi^{3/2} v_{\rm th}^3 } \exp 
\biggl [ - {(v_\parallel-(R B_t/B)\omega_\phi )^2 + 2 \mu B / m \over v_{\rm th}^2} \biggr ] , 
\ee 
which determines the source term,
\bee
S =  - ({\bf v}_\chi) \cdot \biggl [ {\nabla n \over n} + \biggl ( {v_\parallel^2 \over v_{\rm th}^2 } 
+ {(\mu B) \over T} - {3 \over 2} \biggr ) {\nabla T \over T} \\
 + {m v_\parallel \over T} {R B_t \over B }\nabla \omega_\phi \biggr ] F_M -  
{Ze \over T} [ v_\parallel {\bf b} + {\bf v}_D ] \cdot \nabla \langle \phi \rangle  F_M.
\label{source} 
\eee
\noindent
The thermal velocity $v_{\rm th}\equiv \sqrt{ 2 T / m}$, and  the major radius ($R$) are use to normalise the length and time scales.
Using standard gyro-kinetic ordering, the length scale of perturbations along the field line ($R \nabla_\parallel  \approx 1$)  are significantly longer than those perpendicular to the field ($R \nabla_\perp \approx 1/ \rho_*$). Here, $\rho_* = \rho_i / R$ is the normalised ion Larmor radius (where $\rho_i = m_i v_{th} / e B$ and $v_{th} = \sqrt{2 T_i / m_i}$).

The electrostatic potential is calculated from the gyro-kinetic quasineutrality equation, 
\bee 
\label{Poisson}
 \sum_{s}  Z_{s} e \int\langle f \rangle^{\dag}d^{3}{\bf v} +  \nonumber\\
 \sum_{s}\frac{Z_{s}^{2}e^{2}}{T_{s}}\int (\langle\langle\phi\rangle\rangle^{\dag}-\phi)F_{Ms}({\bf v})d^{3}{\bf v} 
 = 0
\eee
where the first term represents the perturbed charge density and the second the polarisation density, which is only calculated from the local Maxwellian.  The dagger operator represents the conjugate gyro-average operator.  Similarly the parallel vector potential is calculated via Amp\`{e}res law,
\bee
-\nabla^2 A_\parallel + \sum_s {\mu_0 Z_s^2 e^2 \over T_s} \int {\rm d}^3 {\bf v} \, v_\parallel^2 \langle 
\langle A_\parallel \rangle \rangle^\dagger F_{Ms} = \nonumber\\
\sum_s {\mu_0 Z_s e \over T_s } \int {\rm d}^3 {\bf v} 
\, v_\parallel \langle g_s \rangle^\dagger 
\label{Ampere}
\eee

The gyro-average is then calculated as a numerical average over a ring with a fixed radius equal to the Larmor radius using a 32 point interpolation defined by,
\be
\langle G \rangle ({\bf X}) = {1 \over 2 \pi }\oint {\rm d} \alpha \, G({\bf X} + \boldsymbol{ \rho})
\ee
where $\alpha$ is the gyro-angle, ${\bf X}$ is the position of the gyro-center and ${\bf \rho}$ is the gyro-radius. This gyro-average is used in both the evolution equation of the 
distribution function, as well as in the quasineutrality and Amp\'{e}re equations and on both the electron and ion species. The nonlinearity (${\bf v}_\chi\cdot \nabla g$) is neglected in this study and its effects on magnetic island evolution and saturation will be left for a future publication. 

\begin{figure}
\centering
\includegraphics[width=8.3cm,clip]{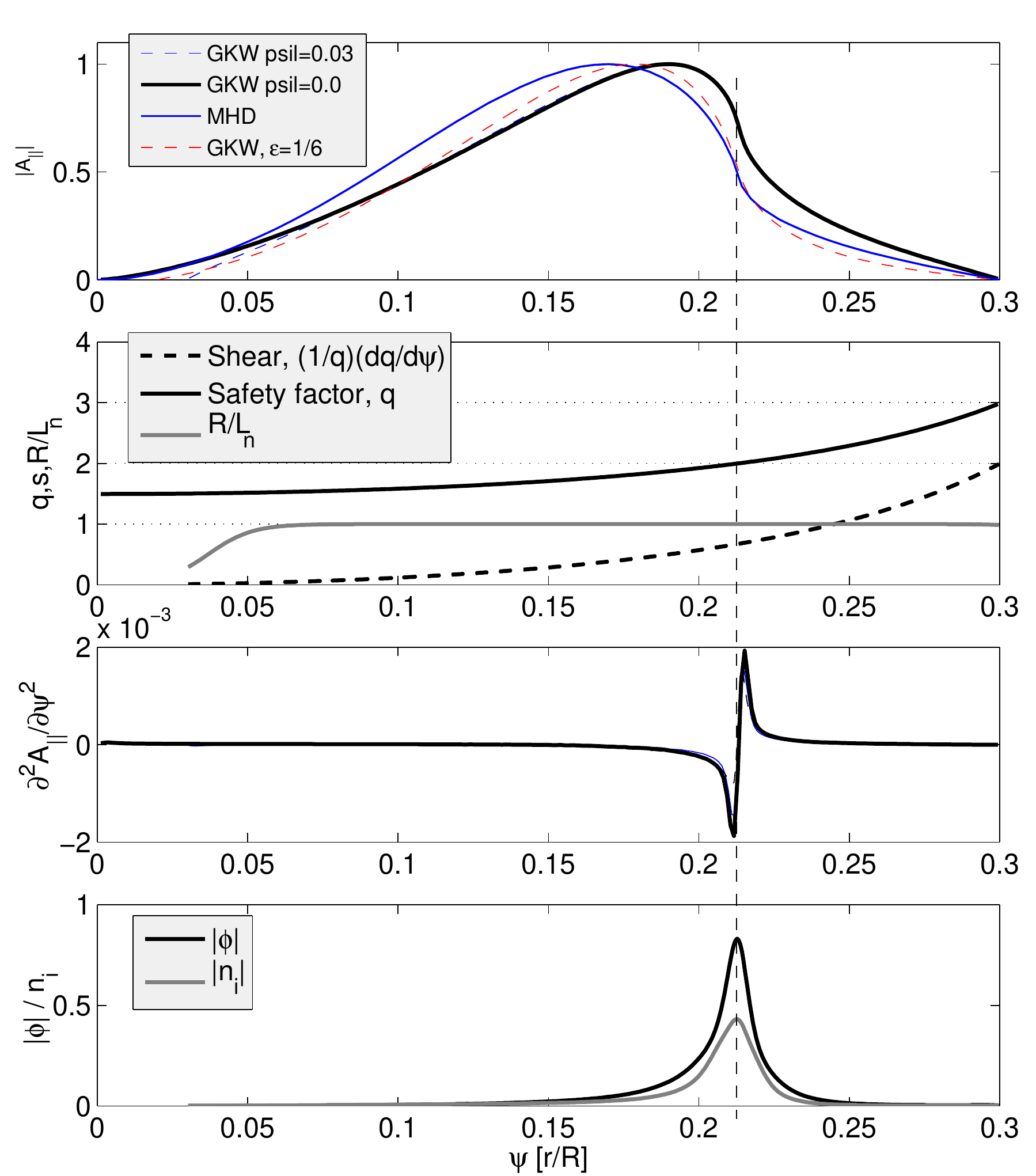}
\caption{The radial profiles of (Top) the parallel vector potential as calculated by
GKW and the ideal MHD solution in cylindrical geometry calculated using the shooting 
method. The shift in the eigenmode is compatible with previous results when comparing the cyclindrical to toroidal
eigenfunction.\cite{NISH98}  (Middle top) The safety factor (q),magnetic shear ($\hat{s}$) and, when present, density gradient ($R/L_{n}$) profiles (Middle bottom)
the radial second derivative, $\partial^{2}A_{||}/\partial\psi^{2}$, indicating the position of the resistive
layer.   (Bottom) are the electrostatic potential ($\phi$) and perturbed ion density ($n_{i}$) profiles.  }
\label{2-1MHD}
\end{figure}

GKW uses straight field line Hamada \cite{HAM58} coordinates ($s,\zeta,\psi$) where $s$ is the coordinate along the magnetic field and $\zeta$ is the generalised toroidal angle and $\psi = r/R_{ref}$ is the normalised radial coordinate. For circular concentric surfaces \cite{LAPcirc}, the transformation of poloidal and 
toroidal angle to these coordinates is given by \cite{PEE09} ($s,\zeta) = (\theta / 2 \pi , [q \theta - \phi]/ 2 \pi)$.
The wave vector of the island is 
\begin{math}
k_\zeta^{I} \rho_i = 2 \pi n \rho_*,
\end{math}
where $n$ is the toroidal mode number.  GKW uses a Fourier representation in the binormal direction, perpendicular to the magnetic field. The radial and parallel directions are treated using a high order finite difference scheme so as to include global profile effects utilising Dirichelet boundary conditions in the radial direction, and open field line boundaries in the parallel direction.  An example of the radial profiles used is seen in the middle panel of Fig.~\ref{2-1MHD}.

The collision operator is particularly important to the growth and development of the tearing mode, as it can be resistive in nature.  In GKW the full
linearised Landau collision operator with momentum and energy conserving terms is implemented.   However here, unless stated otherwise, we utilise only the pitch-angle scattering part.  The differential part of the operator has the following form \cite{KAR86} 
\begin{eqnarray}
C(f_a) = \sum_{b}
{1 \over v 
\sin \theta } {\partial \over \partial \theta} \biggl [ \sin \theta  D_{\theta \theta}^{a/b} 
{1\over v}{\partial {f_a} \over \partial \theta} \biggr ] .
\end{eqnarray}
Here the sum is over all species $b$ and $\theta$ denotes the particle pitch-angle ($v_{||}=v\cos{\theta}$).

The coefficients can be obtained from the literature \cite{KAR86}  and written in normalised form, suppressing the subscript N for the normalised quantities:	
\be 
D_{\theta \theta}^{a/b} = \sum_{b} {\Gamma^{a/b} \over 4 v_{a}} \biggl [ \biggl ( 2 - {1\over v_{b}^2} \biggr ) 
{\rm erf}(v_{b}) + {1 \over v_{b}} {\rm erf}^\prime (v_{b}) \biggr ] ,
\ee
where ${\rm erf}$ and ${\rm erf'}$ are the standard definition of the error function and its derivative.  Finally, the normalised collision frequency, $\Gamma^{a/b}$ is given by 
 
$\Gamma^{a/b} = {R_{\rm ref} n_b Z_a^2 Z_b^2 e^4 \ln \Lambda^{a/b} \over 4 \pi \epsilon_0^2 m_a^2 v_{tha}^4}$,
 $n_{b}$ is the scattering species number density, Z is the relative charge of the species and $\ln\Lambda$ is the Coulomb logarithm. $v_{a}$ and $v_{b}$ are the velocities of the scattered and scattering species respectively.

\section{Tearing mode implementation}
\label{Implementation}

The tearing mode instability is driven by a non-homogeneous current density.  In our implementation this 
is introduced by applying an electron flow profile in the equilibrium.   The electron flow velocity is calculated self-consistently from the imposed q-profile  analogous to the method used in \cite{DRA77} for a kinetic calculation in slab geometry. 

We assume that the background current is carried by the electrons, $J = -n_{e}v_{e}e$, i.e. that the electron flow velocity $v_{e}$ is larger than the ion flow and that it is a flux function.  The current gradient is 
therefore related to the gradient in the electron flow, $\partial u_{e}/\partial\psi$, this profile is introduced to the code where $\nabla\omega_{\phi}$ is 
written in equation \ref{source}. 

We assume $v_{e}$ to be significantly smaller than the electron thermal velocity, thus the only term in Eq.~\ref{source} of interest is,
\begin{math}
\nabla F_{Me} = -2 \frac{v_{||}}{v_{th}} \nabla u_{e} F_{Me}.
\end{math}

The electron profile is related to the q-profile via Amp\'{e}res law, $\int\vec{B}\cdot ds = \mu_{0}\int j d^{2}A$,
which in circular geometry and assuming a low aspect ratio, becomes,
\begin{equation}
2\pi r \mu_{0} B_{p0}F(r) = 2\pi\mu_{0}\int_{0}^{r} r'dr' j(r'),
\end{equation}
where the function $F(r) = 1/2\pi\int d\theta/(1+\epsilon\cos{\theta})$ is approximated to be close to unity and subscript zeros represent constant values.
We have used the definition $q= \frac{1}{2\pi}\int\frac{rB_{t}}{RB_{p}} = \frac{rB_{t0}}{R_{0}B_{p0}}F(r)$ and therefore
the poloidal magnetic field, $B_{p0} = \frac{B_{t0}}{R_{0}}F(r)\frac{r}{q}$.  The gradient of the background current is given by the expression,
\begin{equation}
\frac{dj}{d\psi} = \frac{B_{t0}}{\mu_{0}R_{0}}\frac{d}{d\psi}\left[ \frac{1}{\psi} \frac{d}{d\psi}\left[\frac{\psi^{2}}{q}\right]\right]
\end{equation} 
In this paper we make use of the Wesson analytic form of the current profile \cite{Wess} which is defined by the expression,  
\begin{equation}
j = j_{0}\left(1-\psi^{2}\left(\frac{R}{a}\right)^{2}\right)^{\nu}
\end{equation}
which, in turn, gives a $q$-profile of the form,
\begin{equation}
q = q_{a}\frac{r^{2}/a^{2}}{1 - (1 - r^{2}/a^{2})^{\nu+1}}
\end{equation}
where $q_{a}$ is the safety factor at $\psi = a/R$.  This is related to the $q$ on the axis, $q_{0}$,
by $q_{a}/q_{0} = \nu + 1$, where $\nu$ is an integer that determines that peaking of the current
gradient.

The density and temperature profiles have the radial form,
\begin{eqnarray}
\frac{\partial n_{s}}{\partial \psi} = \frac{1}{2}\frac{R}{L_{ns}}(\tanh{(x-x_{0}+\Delta x)/w}-\nonumber\\ \tanh{(x-x_{0}+\Delta x)/w})\nonumber\\
\frac{\partial T_{s}}{\partial \psi} = \frac{1}{2}\frac{R}{L_{Ts}}(\tanh{(x-x_{0}+\Delta x)/w}-\nonumber\\ \tanh{(x-x_{0}+\Delta x)/w})\nonumber
\end{eqnarray}
where $R/L_{n} = -(R/n)\partial n/\partial\psi$ and $R/L_{T} = -(R/T)\partial T/\partial\psi$ are the logarithmic density and temperature gradients at the reference radius, $x_{0}$, $\Delta x$ is the profile width and $w$ is the rising width.   

\section{Parameters}
\label{Parameters}

In this section we study the linear mode structure, a two-dimensional example of which is plotted in Fig.~\ref{ModeStruct}, showing the radially elognated eigenfunction of the parallel vector potential (left panel) and the electrostatic potential ($\phi$) which is highly localised around the singular layer at the $q=2$ rational surface (dashed line).  The growth rates and mode rotation frequencies are calculated for the following parameters unless stated otherwise in the text:
\begin{center}
\begin{tabular}{| l | l |}
\hline
  \multicolumn{2}{|c|}{Simulation parameters.} \\
\hline
Aspect ratio, {R}/{a} & 3 \\
Electron beta, $\beta_{e}$ & $10^{-3} (0.1\%)$ \\
Current peaking parameter, $\nu$ & 1 \\
Mass ratio, $m_{i}/m_{e}$ & 1836 \\
q at edge, $q_{a}$ & 2.99 \\
Mode number, $k_{\zeta}\rho_{i}$  & 0.0255 \\
$\rho_{*} = \rho_{i}/R$ & 0.002815 \\
$T_{i} = T_{e}$ & \\
$\max(v_{||}/\mu)$ & 4/8\\
\hline
\end{tabular}
\end{center}

Care must be taken in resolving the relevant scale lengths in each of the domain directions.  The number of grid points in the poloidal, parallel velocity and $\mu$ directions were, $N_{s}=32$, $N_{vp}=64$, $N_{\mu}=16$ respectively which were found to give converged growth rates to within a couple of percent.  A single toroidal mode is considered, namely with the $n=1$ mode number, the number of resonant poloidal
modes within the domain is determined by the $q$-profile ($q=m/n$); as such the possibilty of mode coupling effects remain, particularly the toroidal coupling of modes with varying poloidal mode numbers (m) which is known to have a destabilizing effect \cite{Car81}.  The radial resolution required is dependent on the physics of the resonant layer.  In the collisionless
tearing mode the relevant scale length is the electron skin depth, $\delta_{e} = c/\omega_{pe}$, where $\omega_{pe}=\sqrt{n_{e}e^{2}/m_{e}\epsilon_{0}}$ is the
electron plasma frequency and $c$ is the speed of light.  The skin depth is,
\begin{math}
\frac{\delta_{e}}{a} = \frac{\rho_{*}}{\sqrt{\frac{m_i}{m_e}}\sqrt{\beta_{e}}}\frac{R}{a}.
\end{math}  
The $\beta_{e}$ is the electron $\beta$ defined by $\beta_{e} = n_{e}T_{e}/(B_{0}^{2}/2\mu_{0})$,  where $n_{e}$ and $T_{e}$ are equilibrium density and temperature and $B_{0}$ is the magnetic field strength at the outboard midplane.  For collisional modes, the resistivity defines the width of the layer.  Unless otherwise stated,
we have used, $N_{x}=512$ radial grid points between $\psi_{l}=0.03$ and $\psi_{h}=0.3$ gives us three grid points within the skin
depth for a normalised gyro-radius of, $\rho_{*} = 0.002815\ (\sim 1/180)$ and a $\beta$ of $0.1\%$, in the collisionless case without a background pressure gradient.  Increasing collisionality and including a background pressure gradient has the effect of increasing the resonant layer width and with it the number of grid points within the layer.
A hydrogen mass ratio plasma is studied due to a slightly more relaxed skin-depth resolution
requirement over a deuterium mass ratio plasma while still maintaining applicability to experimental conditions. 

\begin{figure}[htpb]
\centering
\includegraphics[width=10.5cm,clip]{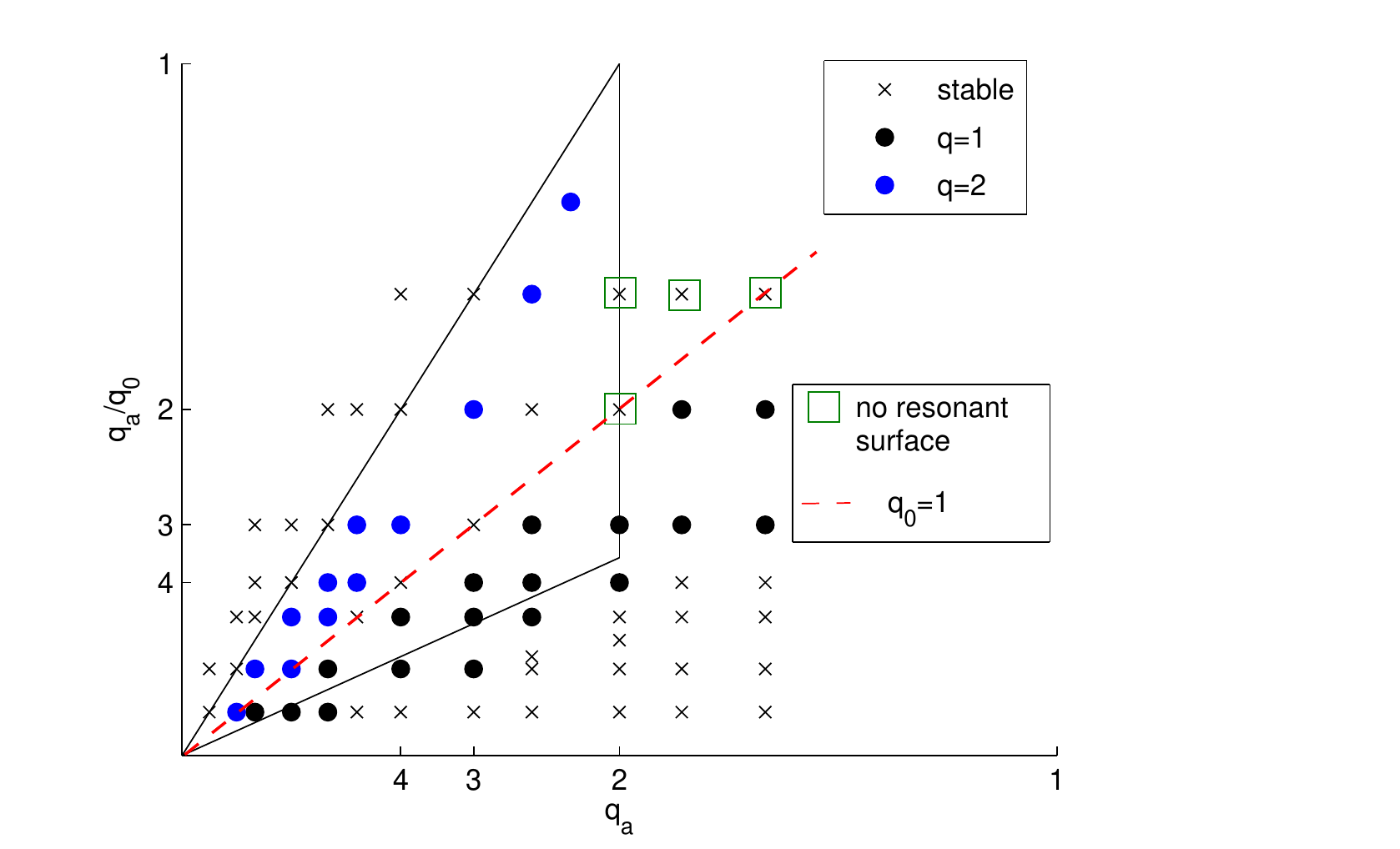}
\includegraphics[width=9.cm,clip]{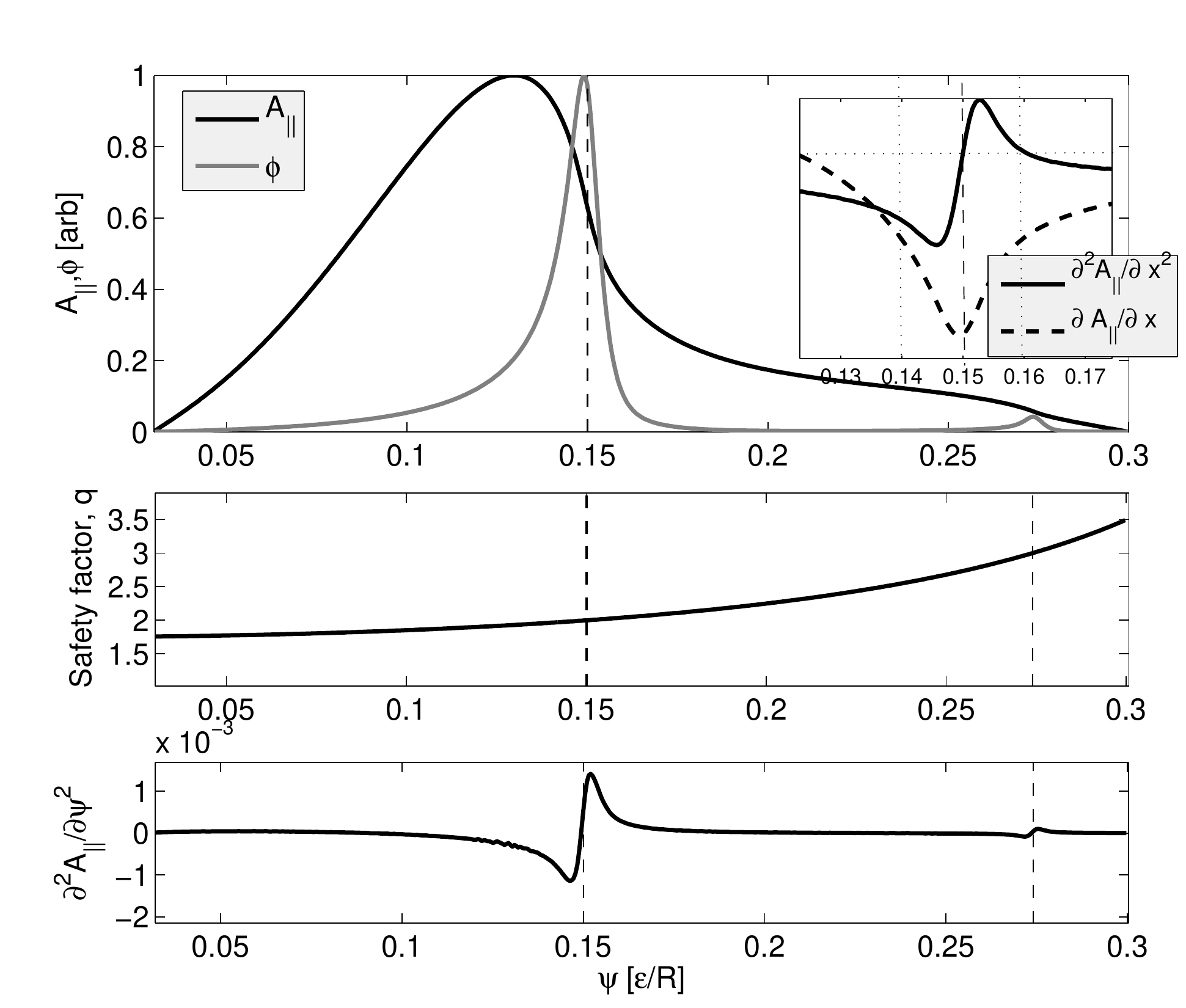}
\caption{(top) Stability of the classical tearing mode varying the current peaking parameter, $\nu = q_{a}/q_{0} - 1$ and the 
safety factor on at the plasma edge, $q_{a}$, normalised collisionality, $\nu_{*}=1.0$.  The lines correspond to the $m=2,n=1$ stability boundaries as found
by Wesson et.al \cite{Wess}.  Solid circles denote an unstable mode, while crosses denote stability. (bottom) The radial mode profiles for a coupled mode with $q_{a}=3.5$ and $\nu=1$.  Inlaid in the top panel are the profiles of the first (dashed line) and second (solid line) radial derivatives  of $A_{||}$ close to the $q=2$ rational surface.  }
\label{lucasfig}
\end{figure}

In Fig.~\ref{2-1MHD} an example of the radial eigenfunctions are plotted and compared with the result from an MHD shooting calculation performed in the cylindrical tokamak limit ($\epsilon\rightarrow 0$).  The safety factor and magnetic shear profiles used in this calculation are found in the second panel.  The shift in the peak of the parallel vector potential with respect to the cylindrical calculation is consistent with previous work performed using an ideal MHD code \cite{NISH98}. Plotted is also the radial eigenfunction for a simulation using an aspect ratio of $R/a = 10$, closer to the cylindrical limit, which shows that the eigenfunction tends toward this cylindrical form as expected.  The radial density and geometry profiles used are shown in middle panel.  The third panel of Fig.~\ref{2-1MHD} shows the second derivative of the radial profile of the vector potential.  This shows clearly the position of the singular layer, the position where ideal MHD breaks down and a discontinuity in the derivative is produced.  The bottom panel also shows the radial profile of the electrostatic potential ($\phi$), showing the highly localised electric field at the rational surface, and the perturbed ion density profile ($n_{i}$).

\section{Tearing mode stability.} 
\label{curprof}

The upper panel of Fig.~\ref{lucasfig} shows the stability of the tearing mode as a function
of the current peaking parameter, $\nu$ and the safety factor at the plasma edge, $q_{a}$.  This is analogous
to the diagram shown in \cite{Has77,Wess}, which shows the stability of the $n=1$, $m=2$ mode in cyclindrical geometry, $n$ denoting the toroidal
mode number, while $m$ is the poloidal mode number.  

The analysis in \cite{Wess} considered a single $m=2$, $n=1$ mode, the outlines of which are also plotted in overlay (black lines). However
we are unable to decompose the modes in this manner and consider, for example a $m=3$, $n=1$ mode in isolation.  The different coloured points denote the dominant mode within a particular simulation determined by the amplitude of the mode at the relevant rational surface.

It is evident that there is good agreement with previous work, particularly when there exists only
an $m=2$, $n=1$ mode within the domain.  However, the results are complicated by the toroidal coupling of modes.  Toroidal coupling has the effect of destabilising modes that would otherwise be stable \cite{Car81,Con87}, for example in lower panel of Fig.~\ref{lucasfig} we see the radial eigenstructure of a combined $m=2,n=1$ and a $m=3,n=1$ mode, the latter of which would otherwise be stable.  The point at $q_{a}=4$, $\nu=2$ has only a $q=3$  rational surface within the domain and is found to be stable.  This is further manifested in the growth rate where this is plotted if Fig.~\ref{lincollscan} for the combined $m=2,3$ mode showing a general increase over the pure $m=2,n=1$ mode. 

The growth rate of the tearing mode is a function of the well known parameter \cite{FUR63},
\begin{equation}
\left.\Delta' = \frac{1}{A_{||}}\frac{\partial A_{||}}{\partial r}\right|^{r_{s}^{+}}_{r_{s}^{-}}
\end{equation}
 across the singular layer, whose position is denoted by $r_{s}$ and upper and lower boundaries by $r_{s}^{+}$ and $r_{s}^{-}$ .  This parameter represents the discontinuity in $\partial A_{||}/\partial r$ and measures whether mode growth is energetically favourable ($\Delta' > 0$ for the mode to grow).  As the growth rate of the mode is directly proportional to $\Delta'$, measuring the growth rate allows us to study the mode stability.   As such the above difference in growth rate between the coupled tearing mode and the pure $m=2, n=1$ shows that the toroidal coupling in this case has a destabilising effect (Increasing $\Delta'$),  increasing the growth rate by a factor of 2.

It can be seen that points along the line $q_{0}=1$ are stable to the tearing mode.  For the collisionality used in the figure ($\nu_* \sim 1.0$, $S=5.10^{4}$), a toroidal MHD analysis (Fig.8 in \cite{Has77}) predicts that points along this line would indeed be stable, and the results from GKW are consistent with this, however, the mode is stabilised at a much lower Lundqvist number, $S$ than is predicted in \cite{Has77}.  $S$ is defined as the ratio of the Alven time ($\tau_{A} = a\sqrt{\mu_0 n_i m_i}/B$) and the resistive time scale $\tau_{R} = \mu_{0}a^{2}/\eta$, where $\eta=m_{e}\nu_{ei}/(n_{e}e^{2})$.

\section{Collisionality effects on growth rate and rotation.}
\label{colleff}

\begin{figure*}[htpb]
\centering
\includegraphics[width=18.5cm,clip]{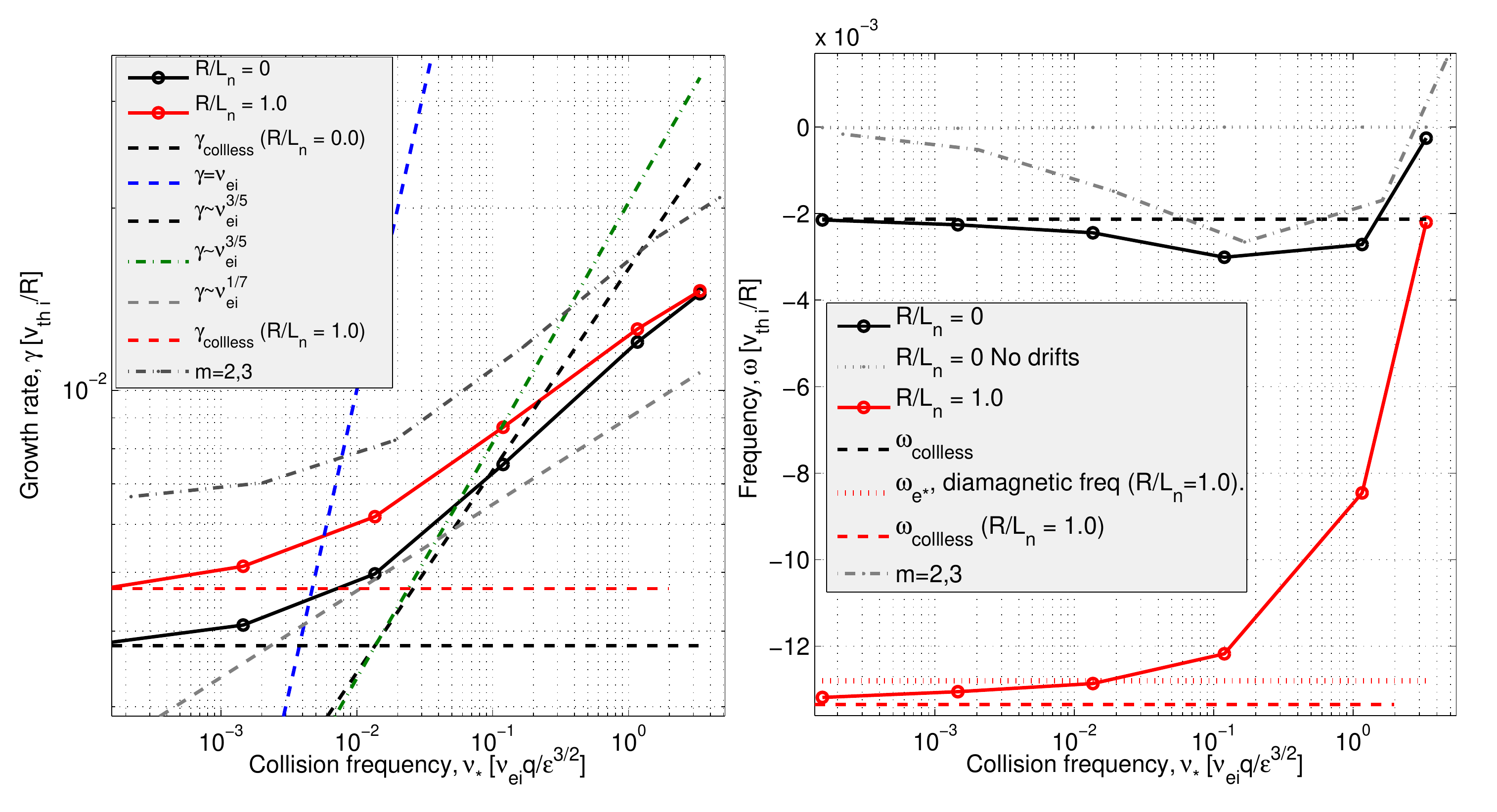}
\caption{The (left) growth rates and (right) mode frequencies of an $m=2,n=1$ tearing mode with aspect ratio $R/a = 3$ in a hydrogen plasma.  These have been
calculated for purely electron-ion pitch angle collision operator with no background pressure gradient (black circles)
and with a background density gradient (red circles).  The left hand panel also shows the analytical result derived by Drake and Lee
for a collisional tearing mode ($\gamma\propto\nu_{ei}^{3/5}$, green dashes) and semi-collisional mode($\gamma\propto\nu_{ei}^{1/3}$,black dashed line).  }
\label{lincollscan}
\end{figure*}

Fig.~\ref{lincollscan} shows the growth rate of the tearing mode as a function of the normalised collisionality (normalised to the trapping-detrapping rate, $\nu_{N} = \nu_{ei}q/\epsilon^{3/2}$), corresponding to a Ludquist number, S, range between $10^{12}$ and $2\times 10^{3}$.  

A scan with no background density gradient (Black circles) and with a background gradient scale length of $R/L_{n} = 1.0$ (Red circles) is presented.   Firstly, at low collisionality the collisionless tearing mode is observed, their growth rates are plotted as horizontal dashed lines which are the low collisionality asymptotic values.  These asymptotic  values were checked for convergence by doubling the radial resolution ($n_{x}=1024$), the growth rate changed by less than 3\% from the value using $n_{x}=512$.  

\begin{figure}
\centering
\includegraphics[width=9.0cm,clip]{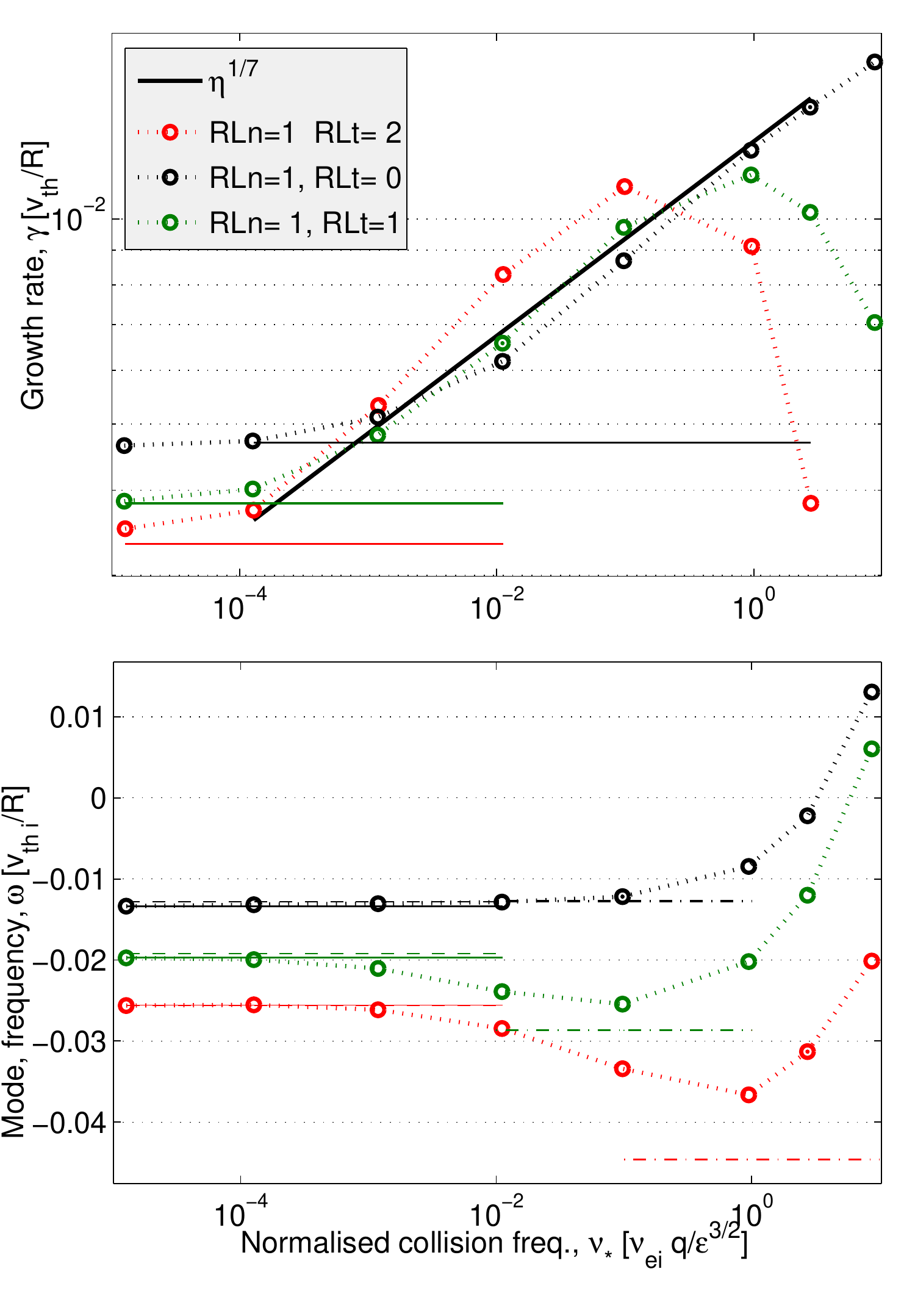}
\caption{The (top) growth rates and (bottom) mode frequencies of an $m=2,n=1$ tearing mode as a function of the normalised collision frequency at the rational surface in the presence of a background temperature and density gradients. Horizontal dashed lines (top) represent the collisionless growth rates, while solid lines (bottom) represent the analytic mode frequencies as calculated by $\omega = \omega_{n*}+0.5 \omega_{T*}$ in the collisionless limit. Dot-dashed lines are the corresponding analytical values for the semi-collisional frequency, $\omega = \omega_{*n} + 5/4\omega_{*T}$.}
\label{lincollscan2}
\end{figure}

As the collisionality is increased (and as such the resistivity) we see the expected increase in the mode growth rate.  The (blue dashed) line represents the condition  $\gamma = \nu_{ei}$ and marks the boundary between the collisionless and semi-collisional mode regimes, at this point collisions limit the electron response, which allows them to be accelerated, and broadens the resonant layer width.  The semi-collisional theory only holds in the limit when $\gamma < \nu_{ei}$ and the transition between semi-collisional and collisionless modes is clearly seen in our simulations.  The growth rate obtained deviates significantly from the collisionless rate once the growth rate is greater than the electron-ion collision frequency.   

Plotted are the analytic scalings for the resistive (Green dashes, $\eta^{3/5}$) and the semi-collisional tearing mode (Black dashes, $\eta^{1/3}$) obtained using a kinetic treatment by Drake and Lee \cite{DRA77}.  However both of these give significantly stronger scalings than our calculated values.  Good agreement, particularly when a density gradient is present, is found when we consider the semi-collisional scaling at high $\Delta'$ as given by the three fluid, two dimensional slab calculation by Fitzpatrick \cite{Fitz10}. This predicts a scaling of the  growth rate with $\eta^{1/7}$ (Light grey dashes).  This corresponds to regime III in Fig.1 of \cite{Fitz10} and was also found in a semi-collisional study of $m=1$ modes \cite{Ade91}.   The visco-resistive scaling as calculated by Porcelli ($\eta^{5/6}$) \cite{Porc87} was also considered, but the scaling is found to be much too strong.

The three fluid theory  \cite{Fitz10} utilises some ordering assumptions in its derivation; firstly that the electron beta is of the order of the mass-ratio ($\sqrt{\beta_{e}}\sim \sqrt{m_{e}/m_{i}}$), secondly we are in the low collisionality limit and finally, that all length scales are larger then the electron gyro-radius, both of which also hold in our calculations.  This model was also derived without a background pressure gradient, however the scaling from our result seems largely invariant to this, as seen in the further two curves in Fig.~\ref{lincollscan2}.   The $\eta^{1/7}$ scaling found here requires a large $\Delta'$ in its derivation.  This requirement, in turn, invalidates the constant-$\psi$ approximation.   To achieve a constant perturbed magnetic flux across the singular layer, diffusive processes must act to smooth out any perturbations, thus the growth rate of the mode must be slower than the time scale of the diffusive process or the magnetic island must be very narrow \cite{Wael93}.   A rough estimate, using the resistive growth rate, states that the growth rate $\tau_{A}\gamma \ll S^{-1/3}$ for the constant-$\psi$ approximation to hold.  This equality is never really satisfied in the present simulations. In Fig~\ref{lucasfig} the radial profile of $\partial A_{||}/\partial\psi$ shows that through the singular layer $A_{||}$ can not be considered constant.  Direct calculation of $\Delta'$ in this case is difficult in toroidal geometry where the resonant layer width is not precisely defined, as such it can be merely estimated qualitatively.  The kinetic calculation of \cite{DRA77} utilises the constant-$\psi$ approximation, as such, with the parameters considered here, a scaling with $\eta^{1/3}$ is not expected (Of note, an assumption of $\Delta'\rightarrow\infty$ is utilised in \cite{Ade91} for an $m=1$ perturbation, where the constant-$\psi$ approximation can not be made).  

The right hand panel of Fig.~\ref{lincollscan} shows the frequency of the tearing mode as a function of the normalised collisionality.   We note that the tearing mode has a small but finite rotation frequency in the absence of a background density or temperature gradient.  Kinetic analysis of the classical tearing mode \cite{DRA77,HAZ75} in slab geometry has shown that, in the absence of equilibrium pressure gradients, the mode has no frequency ($\omega=0$) which is in contrast to our calculated values.  Our calculation is a more complete description, performed in three dimensional geometry, and thus includes trapping and curvature drift effects.  By artificially switching off the terms related to particle drifts in the gyro kinetic equation, it can be seen that this residual frequency is indeed a toroidal effect.  When curvature effects are removed, the mode frequency is found to be essentially zero (Right hand panel of Fig~\ref{lincollscan}).  At high collisionalities, (i.e  $\nu_{*}>1$) the frequency begins to decrease and eventually changes sign, rotating in the ion diamagnetic direction, something not predicted by previous analyses.

\begin{figure}
\centering
\includegraphics[width=9.0cm,clip]{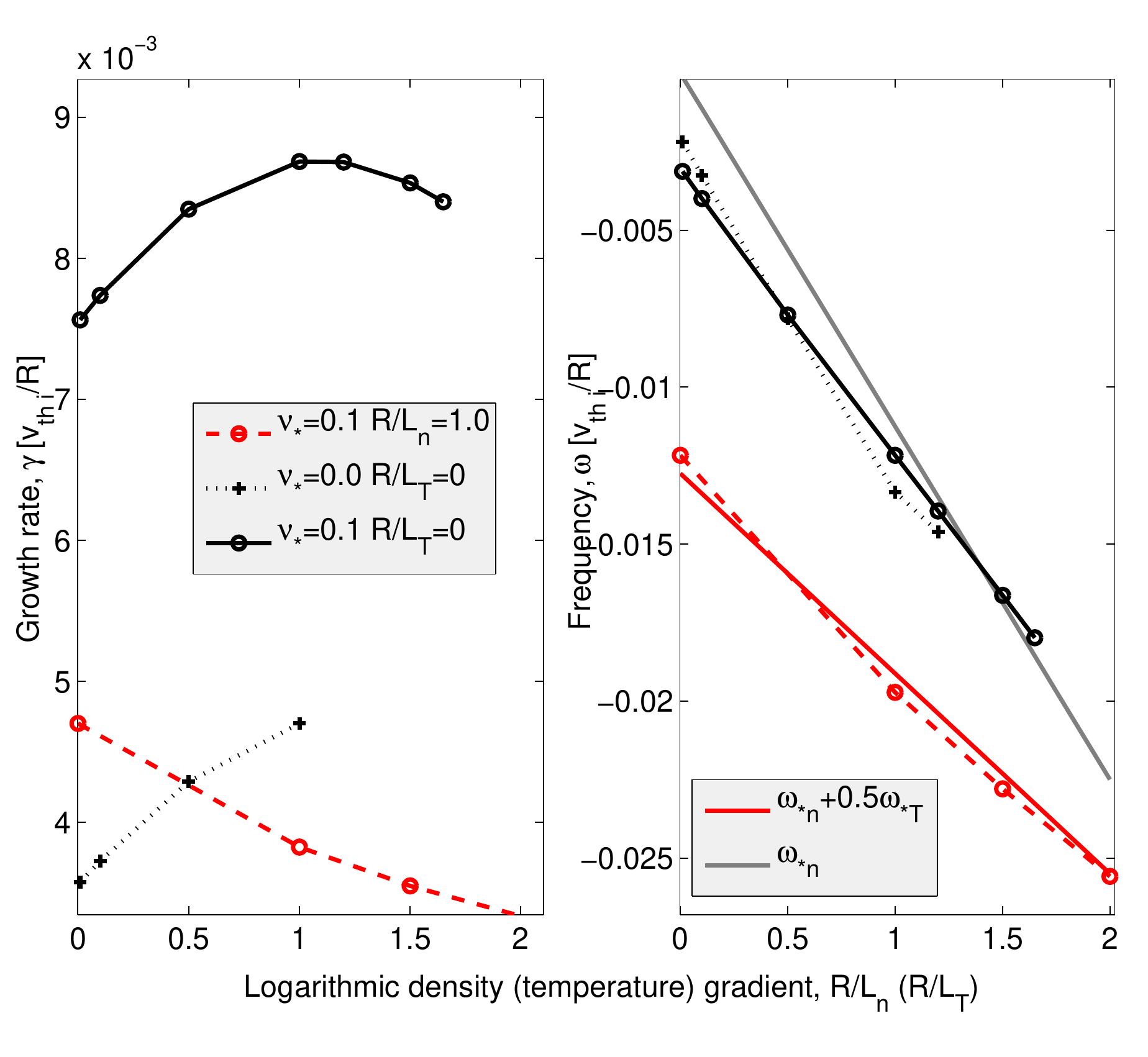}
\caption{The growth rate (left) and frequency (right) of the mode as a function of the equilibrium density gradient at the resonant surface ($R/L_{n}$, black) or equilibrium temperature gradients ($R/L_{T}$, red).  The solid grey line is the electron diamagnetic frequency when considering only $R/L_{n}$ while red is the Drake and Lee prediction of the collisionless mode frequency $\omega_{*n}+0.5\omega_{*T}$.}
\label{dengradfig}
\end{figure}

Presented in Fig.~\ref{lincollscan2} is a collisionality scan (over the same range of parameters as Fig.~\ref{lincollscan}) showing the growth rate and mode frequency in the presence of three different combinations of background density ($R/L_{n}$) and temperature gradients ($R/L_{T}$).  The same scaling is seen as in Fig.~\ref{lincollscan}, however it is apparent that at high collisionality the mode starts to stabilise, apparent by the reduction of the growth rate.

Our result shows that the presence of larger pressure gradients causes a stabilisation of the mode at lower collisionalities.  This is further evident in Fig.~\ref{dengradfig} where the growth rate and mode frequency is shown as a function of the equilibrium density gradient at fixed collisionality.   Initially here the density gradient has a destabilising effect, causing the growth rate to increase.  At higher gradients ($R/L_{n} > 1.0$), the mode begins to stabilize.  There are two key parameters in determining the nature of the tearing instability when considering the effects of equilibrium pressure gradients \cite{Dra83,Con12}, which are noted for having  a stabilising effect on the mode as they produce currents outside the resonant layer.  The currents shield the singular layer from the magnetic perturbation, preventing reconnection and thus stabilising the TM \cite{Dra83}.  This effect is a function of the normalised beta, $\hat{\beta}=\beta_{e}/2(L_{s}/R)^{2}(R/L_{n})^{2}$ where $L_{s}=Rq/\hat{s}$ is the magnetic shear length, and a second parameter, a normalised collisionality, $C=0.51(L_{s}/L_{n})^{2}(\nu_{ei}/\omega_{*e})(m_{e}/m_{i})$ which reduces to $0.51(\delta_{c}/\rho_{i})^{2}$ where  $ \delta_{c}$ is the semi-collisional resonant layer width.   In the simulations presented it is possible that $\delta_{c}\sim\rho_{i}$ (\cite{Con12} assume that $\rho_{i} > \delta_{c}$).  

It has been shown analytically \cite{Dra83} that as the parameter $\hat{\beta}$ approaches unity the tearing mode becomes entirely stable.  The results presented here have values significantly lower, ($\hat{\beta}\sim 0.05(R/L_{n})^{2}$ where $\hat{s}=0.65$ at the $q=2$ rational surface) but the stabilising effects of pressure gradients still become apparent.  Analytical work has shown that toroidal geometry has a stabilising effect on the tearing mode\cite{GGJ75}, however, at a low beta, as is used in this work, these effects will be small

The mode frequency of the tearing mode is plotted in Fig.~\ref{lincollscan2} in the presence of a density and temperature gradient as a function of collisionality along with the corresponding collisionless asymptotes, while in Fig.~\ref{dengradfig} as a function of density/temperature gradient at fixed collisionality.   Kinetic analysis of the mode shows that it rotates at the electron diamagnetic frequency related to the density and temperature gradient at the resonant layer when they are present.  In GKW units the diamagnetic frequency  is given by $\omega_{*e} = -(R/L_{n})(k_{th}\rho_{i})/2 = -0.0128 v_{th i}/R$, for $R/L_{n}=1.0$ and $R/L_{T}=0.0$.  When a background temperature gradient is also considered then, in the collisionless limit, the mode frequency is given by the expression, $\omega = \omega_{*n} + 1/2\omega_{*T}$ \cite{DRA77}, where $\omega_{*T} = -1/2(k_{\theta}\rho_{i})R/L_{T}$ which is the electron diamagnetic frequency associated with the temperature gradient, these prediced values are also plotted and their scalings have been plotted in the right hand panel of Fig.~\ref{dengradfig}.  In both cases showing a good agreement.  We also see that the residual frequency as previously mentioned, is significantly smaller than the electron diamagnetic frequency when the mode evolves in the presence of pressure gradients producing the so called drift-tearing mode.

Plotted in the bottom panel of Fig.~\ref{lincollscan2} are the mode frequencies with (horizontal solid lines) the corresponding diamagnetic frequency as given by the above expression, giving good agreement in the collisionless limit. It is also found that the frequency never reaches the Drake and Lee semi-collisionless expression of,  $\omega = \omega_{*n} + 5/4\omega_{*T}$ which are also plotted as horizontal dot-dashed lines, although a qualitative agreement as a function of the pressure gradient is seen i.e. with a larger background pressure gradient a larger maximal frequency is seen.   It is also noted that the collisionality threshold at which the mode changes the sign of its rotation is a function of the pressure gradient,  with a higher threshold seen at higher background gradients.  While the mode start to be stabilised at a lower collisionality, the opposite relation.

\section{Conclusions}

Using the global version of the gyrokinetic code, GKW, the tearing mode has been studied in the linear regime for experimentally relevant paramerters in three dimensional, toroidal geometry for the first time.  The collisionality, equilibrium temperature, density and current gradients were varied and their effects on the mode stability and frequency were analysed.  The main results can be summarised as follows:

\begin{itemize}

\item The tearing mode growth rate scales with the resistivity close to the $\eta^{1/7}$ scaling as calculated by Fitzpatrick \cite{Fitz10} in the semi-collisional, high $\Delta'$ regime as opposed to the semi-collisional scaling of $\eta^{1/3}$ found in the kinetic treatment of Drake and Lee. 

\item The stability of the mode follows closely the stability calculated in cylindrical geometry performed by Wesson et. al. \cite{Has77}.  However,  our analysis differs in that multiple rational surfaces may be present within the domain and so allows double tearing modes and toroidal coupling of modes, a process which is known to be destabilising, as such more modes are excited in our global toroidal treatment.

\item The mode is seen to rotate at the electron diamagnetic frequency when a pressure gradient is present, as is characteristic for the drift-tearing mode.  However, this is a function of the collisionality and the sign of the rotation frequency is seen to change at a sufficienty high collisionality. 

\item A residual non-zero, mode frequency, is found for the classical tearing mode in three dimensional toroidal geometry, even when equilibrium pressure gradients are not included.  Suppressing toroidal drift effects recovers a zero mode frequency.  

\item Pressure gradient stabilisation of the mode is apparent, even though $\beta_{e}$   which determines the effective stabilisation, is small.

\end{itemize}

Linear mode evolution is valid only when the generated magnetic island is small compared to the resonant layer width, which in these simulations were smaller than the ion gyroradius.  However, the mode frequency is important for the non-linear evolution of the magnetic island and in particular the polarisation current drive of the tearing mode, and its neoclassical modification, the Neoclassical tearing mode, particularly when the magnetic island is small \cite{Pol05}.

\begin{acknowledgments}
A part of this work was carried out using the HELIOS supercomputer system at Computational Simulation Centre of International Fusion Energy Research Centre (IFERC-CSC), Aomori, Japan, under the Broader Approach collaboration 
between Euratom and Japan, implemented by Fusion for Energy and JAEA.

The authors would like to thank the Lorentz Center, Leiden for their support.  Useful and productive conversations with
C.S. Chang, C. Hegna, G. Huismanns and other attendees of the meeting, ``Modelling Kinetic Aspects of Global MHD Modes'' are gratefully acknowledged.

\end{acknowledgments}


\begin{thebibliography}{10}

\bibitem{Bisk} D.~Biskamp {\it Magnetic reconnection in plasmas. 3rd Edition} Cambridge University Press (2005)

\bibitem{NEW60} W.~A.~Newcomb, Annals of Physics {\bf 10} 232-267 (1960)

\bibitem{FUR63} H.~P.~Furth, J.~Killeen, M.~N.~Rosenbluth, Phys. Fluids {\bf 6} 459 (1963)

\bibitem{FUR73} H.~P.~Furth, M.~N.~Rosenbluth, H.~Selberg, Phys. Fluids {\bf 16} 1054 (1973)

\bibitem{Zwei09} E.G.~Zweibel and M.~Yamada, Annu. Rev. Astron. Astrophys. {\bf 47} 291 (2009)

\bibitem{WAE09} F.~L.~Waelbroeck, Nucl. Fusion {\bf 49} 104025 (2009)

\bibitem{CAR86} R. Carreras, R.D. Hazeltine, M. Kotschenreuther, Phys. Fluids {\bf 29} 899 (1986)

\bibitem{HEG98} C.C.~Hegna, Phys. Plasmas {\bf 5} 1767 (1998)

\bibitem{WIL96} H.R. Wilson, J.W. Connor, R.J. Hastie, and C.C. Hegna, Phys. Plasmas {\bf 3} 248 (1996)

\bibitem{SAU97} O.~Sauter {\it et. al}, Phys. Plasmas {\bf 4} 1654 (1997)

\bibitem{NISH98} Y.~Nishimura, J.~D.~Callen, C.~C.~Hegna, Phys. Plasmas {\bf 5} 4292 (1998)


\bibitem{RUTH73} P.~H.~Rutherford, Phys. Fluids {\bf 16} 1903 (1973)

\bibitem{Has04} R.J.~Hastie, F.~Militello and F.~Porcelli, Phys. Rev Lett. {\bf 95} 065001 (2005) 

\bibitem{PRL14} W.~A.~Hornsby et al. (Submitted) arXiv:1403.1520

\bibitem{POL09} E. Poli, A. Bottino and A.G. Peeters, Nucl. Fusion {\bf 49}, 075010 (2009)

\bibitem{HorEPL} W.A. Hornsby et.al., Euro. Phys. Lett. {\bf 91} 45001 (2010)

\bibitem{HAZ75} R.~D.~Hazeltine, D.~Dobrott, T.~S.~Wang, Phys Fluids {\bf 18}, 1778 (1975)

\bibitem{DRA77} J.~F.~Drake, Y.~C.~Lee, Phys. Fluids {\bf 20}, 1341 (1977)

\bibitem{Fitz10} R.~Fitzpatrick, Phys. Plasmas {\bf 17}, 042101 (2010)

\bibitem{Ade91} A.~Y.~Aydemir, Phys. Fluids B. {\bf 3} 3025 (1991)

\bibitem{Wael93} F.L.~Waelbroeck, Phys. Rev. Lett. {\bf 70} 3259 (1993)

\bibitem{Hahm85} T.~S.~Hahm and L.~Chen, Phys. Fluids {\bf 29} 1891 (1986)

\bibitem{Fitz06} R. Fitzpatrick, F. L. Waelbroeck, and F. Militello, Phys. Plasmas, {\bf 13} 122507 (2006)

\bibitem{Pol05} E.~Poli, A.~Bergmann and A.~G.~Peeters, Phys. Rev. Lett. {\bf 94} 205001 (2005)

\bibitem{NUM09} R.~Numata, W.~Dorland, G.~G.~Howes, N.~F.~Loureiro, B.~N.~Rogers, T.~Tatuno, Phys. Plasmas {\bf 18}, 112106 (2011)

\bibitem{ROG07} B.~N.~Rogers, S.~Kobayashi, P.~Ricci, W.~Dorland, J.~Drake {\it et al.}, Phys. Plasmas {\bf 14}, 092110 (2007)

\bibitem{Wan05} W.~Wan, Y.~Chen, S.~E.~Parker, Phys. Plasmas {\bf 12} 012311 (2005)

\bibitem{Mish} A.~Mishchenko and A.~Zocco, Phys. Plasmas {\bf 19}, 122104 (2012)

\bibitem{PEEGlo} A.G. Peeters et al (To be submitted)

\bibitem{PEE09} A.G. Peeters, Y. Camenen, F.J. Casson, W.A. Hornsby, A.P. Snodin, D. Strintzi, and G. Szepesi, Comp. Phys. Comm. {\bf 180}, 2649 (2009) 

\bibitem{LAPcirc} X. Lapillonne, S. Brunner, T. Dannert, S. Jolliet, A. Marinoni, L. Villard, T. Gorler, F. Jenko, and F. Merz,  Phys. Plasmas, {\bf 16}, 032308, (2009). 

\bibitem{HAM58} S. Hamada, Kakuyugo Kenkyu {\bf 1}, 542 (1958)

\bibitem{KAR86} C.F.F. Karney, Comp. Phys. Reports {\bf 4} (1986) 183

\bibitem{FITZ05} R.~Fitzpatrick, P.~G.~Watson and  F.~L.~Waelbroeck, Phys. Plasmas {\bf 12} 082510 (2005)

\bibitem{Kat80} I.~Katanuma, T.~Kamimura, Phys. Fluids {\bf 23} 2500 (1980) 

\bibitem{Wess} J.~Wesson, \textit{Tokamaks} (Cambridge University Press, Cambridge, 1987)

\bibitem{Porc87} F.~Porcelli, Phys. Fluids {\bf 30} 6 (1987)

\bibitem{Car81} B.~Carreras, H.~R.~Hicks and D.~K.~Lee, Phys Fluids {\bf 24} 1 (1981)

\bibitem{Con87} J.W.Connor, S.C.Cowley, R.J. Hastie, T.C. Hender, A.Hood and T.J.Martin, Phys. Fluids {\bf 31} 577 (1988)

\bibitem{Dra83} J.~F.~Drake, T.~M.~Antonsen Jr., A.~B.~Hassam and N.~T.~Gladd, Phys. Fluids {\bf 26} 2509 (1983)

\bibitem{GGJ75} A.H. Glasser, J.M. Greene and J.L. Johnson, Phys. Fluids {\bf 18} 875 (1975)

\bibitem{Has77} R.J.~Hastie, A.~Sykes, M.~Turner and J.A.~Wesson, Nucl. Fusion, {\bf 17} 3 (1977) 

\bibitem{Con12} J.W.~Connor, R.J.~Hastie and A.~Zocco, Plasma. Phys. Contol. Fusion {\bf 54} 035003 (2012)


\end{thebibliography}
\end{document}